\documentclass[pra,twocolumn,showpacs,floatfix]{revtex4}

\usepackage{graphicx}
\usepackage{amsmath}
\usepackage{amssymb}
\usepackage{epsf,latexsym}
\usepackage{bm}

\usepackage{float}

\newcommand{\ket}[1]{\left| #1 \right\rangle}

\newcommand{\be}{\begin{equation}}
\newcommand{\ee}{\end{equation}}
\newcommand{\ba}{\begin{eqnarray}}
\newcommand{\ea}{\end{eqnarray}}

\begin{document}
\title{Feasibility of approximating spatial and local entanglement in long-range interacting systems using the extended Hubbard model}
\author{J. P. Coe$^{1}$
}
\email{jpc503@york.ac.uk}
\author{V. V. Fran\c{c}a$^{2}$
}
\email{vivian.franca@physik.uni-freiburg.de}
\author{I. D'Amico$^{1}$
}
\email{ida500@york.ac.uk}

\affiliation{                  
  $^1$ The Department of Physics, The University of York, Heslington, York, UK.\\
  $^2$ Physikalisches Institut, Albert-Ludwigs-Universit\"{a}t, Hermann-Herder-Stra\ss e 3, D-79104 Freiburg, Germany.
}

\begin{abstract}
We investigate the extended Hubbard model as an approximation to the local and spatial entanglement of a one-dimensional chain of nanostructures where the particles interact via a long range interaction represented by a `soft' Coulomb potential. In the process we design a protocol to calculate the particle-particle spatial entanglement for the Hubbard model and show that, in striking contrast with the loss of spatial degrees of freedom, the predictions are reasonably accurate. We also compare results for the local entanglement with previous results found using a contact interaction \cite{Coe-Viv-DAm}  and show that while the extended Hubbard model recovers a better agreement with the entanglement of a long-range interacting system, there remain realistic parameter regions where it fails to predict the quantitative and qualitative behaviour of the entanglement in the nanostructure system.
\end{abstract}


\maketitle

\section{Introduction}
 
The properties of quantum dot-based nanostructures make them possible candidates as hardware for the manipulation of quantum information \cite{PhysRevA.57.120,hodgson:114319,FENG04,XIAOQIN03,Pazy03}.  
Chains of quantum dots (QDs) have been proposed to transfer quantum information \cite{damico:2006}, and to generate, distribute, and freeze entanglement \cite{SPILLER07,damico:2007}.
Entanglement is in fact considered a fundamental resource for quantum information processes and is increasingly gaining the attention of the condensed-matter community as a probe for delicate phenomena, such as quantum phase transitions \cite{RevModPhys.80.517}.  

The Hubbard model \cite{OrigHubbard} is a simplified model of itinerant interacting fermions with positions discretized to a lattice, and usually only includes interaction within a lattice site, while extended Hubbard models may include long-range interactions. The calculation of the entanglement in a realistic system of many fermions, such as electrons trapped in a nanostructure, is usually computationally too demanding.  In this respect, proving the Hubbard model -- or a variant of the Hubbard model -- accurate as an approximation to the entanglement of a realistic many-fermion system
would open the possibility of using density-functional theory techniques to calculate the entanglement of these complex systems:  a powerful local-density approximation approach is in fact available for calculating the entanglement of the Hubbard model \cite{VIVIAN1}.

In previous work \cite{Coe-Viv-DAm} the accuracy of the 1D Hubbard model (HM) as an approximation to the average local entanglement  of two-electrons trapped in a chain of quantum dots (QDs) was considered. In \cite{Coe-Viv-DAm} we considered a {\it contact} interaction between particles and found the HM to be accurate for single-site entanglement calculations, but the spatial entanglement could not be estimated as there was no scheme for such entanglement measurement in the HM.  

In this paper we consider a similar system but with a more realistic long range particle-particle interaction, which we model as a `soft' Coulomb potential.  The accuracy of the local entanglement of the 1D extended Hubbard model (EHM) \cite{EHMentanglement1,EHMentanglement2} and the EHM with correlated hopping (EHM+CH) as an approximation to that of the QD structure is then appraised. In addition we propose a method to calculate the spatial entanglement of the HM and EHM and compare this somewhat severe approximation with the spatial entanglement of the QD system. Surprisingly our results show that even if in the HM and EHM cases the number of spatial degrees of freedom are reduced to just a few lattice sites, the HM and EHM spatial entanglement still capture most of the features and trends of the spatial entanglement of the quantum dot structure, and it does so for both attractive and repulsive interactions. 

\section{The QD-chain two-particle system}
\label{sec:Model}
We consider a system of two fermions trapped within a one-dimensional QD-chain. 
In effective atomic units, the Hamiltonian is 
\begin{equation}
H_{QD}=\sum_{s=a,b}\left(-\frac{1}{2}\frac{d^{2}}{dx_{s}^{2}}+v(x_{s}) \right )+C_{U}f(\left|x_{a}-x_{b}\right|).
\label{H_wells}
\end{equation}
Here $v(x_{s})$ is the potential used to represent an array of regularly spaced, identical square wells, each well representing a QD. The chain is symmetric around the origin, and defined by the quantities:  $M$ the number of wells, $d$ the barrier width between two consecutive wells, and $w$ and $v_{0}$ the width and depth of each well respectively.  We define and vary the interaction strength $C_{U}$ to facilitate comparison of the QD system with the HM. The interaction type we consider is either a contact interaction, $f(\left|x_{a}-x_{b}\right|)=\delta(x_{a}-x_{b})$, or a long range interaction of the form (see {\it e.g.} refs.~\cite{Eberly88,Shirwan09})
\begin{equation}
f(\left|x_{a}-x_{b}\right|)=\frac{1}{\sqrt{(l^{2}+(x_{a}-x_{b})^{2})}}.
\label{eq:longrange}
\end{equation}
This is often referred to as a `soft' Coulomb potential. Here we use $l=1$ $a_0$, $a_0$ the effective Bohr radius.  In this work we focus on a system of four wells as previous results \cite{Coe-Viv-DAm} showed that this was the smallest number for which the average local entanglement exhibits a non-trivial dependence on $C_{U}$.  We calculate the solution of the time-independent Schr\"{o}dinger equation corresponding to eq.~(\ref{H_wells}) by using `exact' diagonalization with a basis formed by the eigenfunctions of the non-interacting system.  We note that the system ground-state is a singlet due to the choice of zero magnetisation.

\section{Average single-site (or local) entanglement}Here we consider the average single-site (or local) entanglement of the system ground state. This type of entanglement is relevant for systems of indistinguishable fermions \cite{ZANARDI,IdentialParticlesArxiv}. To this aim we divide our QD system into contiguous `sites', each site centred around a single well.  The entanglement entropy $S$ of the system is given by 
\begin{equation}
S=\frac{1}{M}\sum_{i}^{M}S_{i}, 
\label{eq:avsinglesite}
\end{equation}
with $S_{i}=-Tr \rho_{\text{red},i} \log_{2} \rho_{\text{red},i}$ the {\it i}-site von Neumann entropy of the reduced density matrix $\rho_{\text{red},i}$. By dividing the system into sites and moving to a site-occupation basis the reduced density matrix becomes the $4\times 4$ diagonal matrix \cite{ZANARDI,Larsson05} $\rho_{\text{red},i}=\text{diag}\left[P_i(\uparrow\downarrow),P_i(\uparrow),P_i(\downarrow),P_i(0)\right],$with $P_i(\alpha)$ the probability of double ($\alpha=\uparrow\downarrow$), single ($\alpha=\uparrow$ or $\downarrow$), or zero ($\alpha=0$) electronic occupation at site $i$.   

We calculate the ground-state wavefunction, for an even number $M$ of wells, and from that obtain the occupation probabilities, as described in detail in ref.~\cite{Coe-Viv-DAm}.

\section{Comparison with Hubbard model variants}
\label{sec:deltaHubbard}
We consider the Hamiltonian
\begin{eqnarray}
\nonumber H&=&\sum_{i,\sigma}\left (\hat{c}_{i,\sigma}^{\dagger}\hat{c}_{i+1,\sigma}+h.c\right)\left[\tilde{t}'\left(\hat{n}_{i,-\sigma}+\hat{n}_{i+1,-\sigma}\right)-t\right]\\
&&+\tilde{U}\sum_{i}\hat{n}_{i,\uparrow}\hat{n}_{i,\downarrow}+\tilde{U'}\sum_{i,\sigma,\sigma'}\hat{n}_{i,\sigma}\hat{n}_{i+1,\sigma'},\label{eqn:HubbardHamiltonian}
\end{eqnarray}
where $\tilde{U}$ and $\tilde{U'}$ are respectively the on-site and inter-sites interaction strength, $t$ is the hopping parameter and $\tilde{t}'$ is the correlated hopping term.  $\hat{c}_{i,\sigma}^{\dagger}$ ($\hat{c}_{i,\sigma}$)  are the creation (annihilation) operators for a fermionic particle of spin $\sigma$ at site $i$, and $\hat{n}_{i,\sigma}=\hat{c}_{i,\sigma}^{\dagger}\hat{c}_{i,\sigma}$ is  the particle number operator. To solve eq.~(\ref{eqn:HubbardHamiltonian}) we use exact diagonalization in the single-site occupation basis $\{\ket{\uparrow \downarrow},\ket{\uparrow},\ket{\downarrow}, \ket{0}\}$ with open boundary conditions and an average particle density of $n=n_{\downarrow}+n_{\uparrow}=2/M$. Here $n_{\sigma}$ is the average density of the $\sigma$-spin component. 
Again we calculate the average single-site entanglement according to eq.~(\ref{eq:avsinglesite}) \cite{PhysRevLett.93.086402,VIVIAN,Larsson05}. We note that for $\tilde{t}'=0$, eq.(\ref{eqn:HubbardHamiltonian}) describes the EHM while for $\tilde{U'}=\tilde{t}'=0$ , the HM is recovered.
 
	In order to calculate the equivalent of $t$ for the QD system, $t_{QD}$, we use the same procedure employed in \cite{Coe-Viv-DAm} for the contact interaction. The corresponding on-site, inter-site and correlated hopping for the QD model are calculated from
\begin{equation}
I_{hijk}=\frac{C_{U}}{2}\int \frac{\phi_{h}(x_{a})\phi_{i}(x_{b})\phi_{j}(x_{a})\phi_{k}(x_{b})}{\sqrt{(l^{2}+(x_{a}-x_{b})^{2})}}  dx_{a} dx_{b},
\end{equation}
with $\tilde{U}_{QD}=I_{LLLL}$, $\tilde{U}'_{QD}=I_{LRLR}$, $\tilde{t}'_{QD}=I_{LLLR}$ and $\phi_{L(R)}$ the single-particle ground state of the finite single square well potential, but positioned in the left ($\phi_{L}$) or right ($\phi_{R}$) well.

Usually the hopping parameter $t$ is used to rescale $\tilde{U}$, $\tilde{U'}$ and $\tilde{t'}$, giving the dimensionless interactions $U=\tilde{U}/t$, $U'=\tilde{U'}/t$ and $t'=\tilde{t}'/t$.  For the QD system we obtain $U=\tilde{U}_{QD}/t_{QD}=17865C_{U}$, $U'= 0.28U$ and $t'=2 \times 10^{-5}U$ for $d=w=2$ $a_0$, and $v_0=10$ Hartree. 

\begin{figure}[t!]\centering
  \includegraphics[width=.4\textwidth]{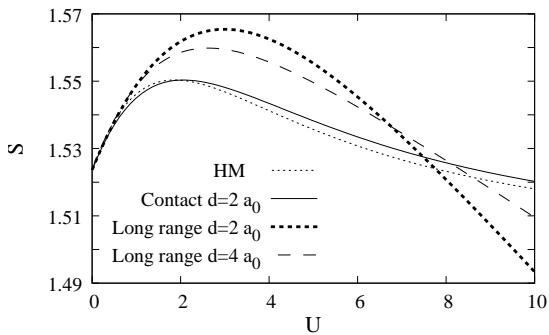}
  \caption{Average single-site entanglement vs interaction strength $U$ for $M=4$. HM (dotted line) and QD system with: contact interaction, $U=12344C_U$ and $d=w=2~a_0$ (solid line); long-range interaction, $U=17865C_U$ and $d=w=2~a_0$ (thick dashed line); long-range interaction, $U=8.5\times10^{7}C_{U}$, $d=4~a_0$ and $w=2~a_0$ (thin dashed line). For all QD systems $v_0=10$ Hartree.}\label{fig:4wellscomparedeltatolongrangeandhubbard}
\end{figure}

We first consider the HM. Fig.~\ref{fig:4wellscomparedeltatolongrangeandhubbard} shows that the HM is not as good an approximation when long range interactions are used. In the QD system with $d=2a_0$, the `long range' average single-site entanglement for $0 < U \lesssim 10$ has some of its behaviour captured by the HM but the long range interaction induces a higher maximum which occurs at a different position, and there is no flex point in the curve.  The calculation of $\tilde{U}_{QD}$ and $t_{QD}$ using the single-particle single square well ground-state may be expected to be a more severe approximation for the long range than for the contact interaction case.  This does not appear to be the main reason for the difference shown in fig.~\ref{fig:4wellscomparedeltatolongrangeandhubbard} though, as a scaling of $U$ by fitting $\tilde{U}_{QD}/(t_{QD}C_{U})$ cannot rectify this difference.  This suggests that there are significant contributions to the interaction beyond the on-site repulsion.

We also investigate the long range interaction case when the wells are further apart ($d=4a_{0}$): this would be expected to reduce the electron density in the barrier region, and make the sites better defined and hence the system more similar to the HM.  We note that our method for calculating $t_{QD}$ becomes in this case too susceptible to the noise in the tail of the numerical wavefunction, as there is an almost negligible overlap between the single-particle wavefunctions centred in the right and left well. So this method fails to give a reliable $t_{QD}$ value.  Hence for $d=4$ $a_{0}$ we use $t_{QD}$ as a parameter to generate a QD entanglement curve closer to the one representing the HM system. With this procedure, the average single-site entanglement's maximum is closer to that of the HM than for $d=2$ $a_{0}$, but the overall shape of the entanglement curve resembles the $d=2$ $a_{0}$ case, as shown in fig.~\ref{fig:4wellscomparedeltatolongrangeandhubbard}. It therefore strongly suggests that much of the dissimilarity with the HM is caused by the long range interaction rather than the proximity of the wells.  In fact,  unlike a contact interaction, a long range interaction does not require the particles' density to overlap.  To test this hypothesis we consider next the EHM and EHM+CH in which interactions between particles on neighbouring sites are included. 

\begin{figure}[ht]\centering
  \includegraphics[width=.4\textwidth]{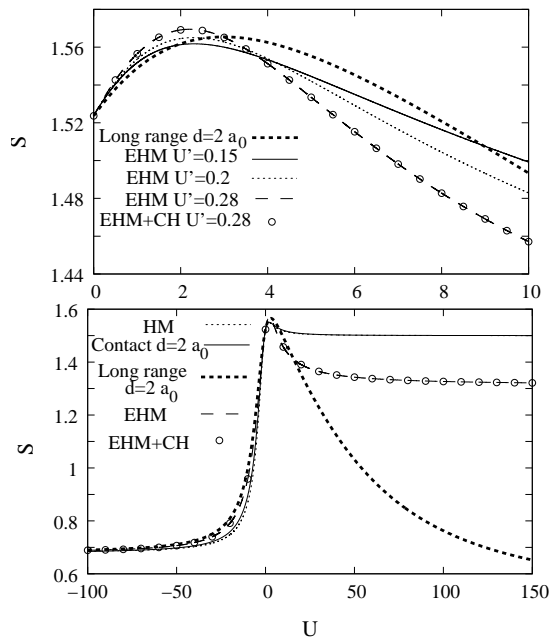}
  \caption{Average single-site entanglement vs interaction strength $U$ for $M=4$. Upper panel: long range interaction, $d=w=2$ $a_{0}$, $U=17865C_{U}$ (thick dashed line); EHM and EHM+CH (circles) with: $U'=0.15U$ (solid line); $U'=0.2U$ (dotted line); $U'=0.28U$ (thin dashed line). Lower panel: contact interaction, $d=w=2$ $a_{0}$, $U=12344C_{U}$ (solid line); HM (dotted line); long-range interaction, $d=w=2$ $a_{0}$, $U=17865C_{U}$ (thick dashed line); EHM (thin dashed line) and EHM+CH (circles) with $U'=0.28U$. For all QD systems $v_0=10$ Hartree.}\label{fig:morepointsUdashcompare}
\end{figure}

For $d=2$ $a_{0}$ the entanglement results for the EHM and EHM+CH are indistinguishable on the scale of the plots (see fig.~\ref{fig:morepointsUdashcompare}).  We see in the upper panel of fig.~\ref{fig:morepointsUdashcompare} that the EHM reproduces better the behaviour of the entanglement when long range interactions are considered,  with the most appropriate $U'$ seemingly residing somewhere between $0.15U$ and $0.2U$.  This difference with our estimated value of $U'=0.28U$ could arise from the use of the non-interacting single square well solutions in our calculation of $U'$.

For attractive interactions, the lower panel of fig.~\ref{fig:morepointsUdashcompare} shows that the EHM reproduces the single-site entanglement of the long-range QD system fairly well at all $U$ values, as well as the HM reproduces the single-site entanglement of the QD with contact interaction.  For repulsive interactions, although the contact interaction and the HM embody some of the features of the long-range interaction system for the parameters chosen, the average single-site entanglement arising from long range interactions is lower for $U \gtrsim 8$, significantly so for very large $U$.  This is due to the long range repulsion forcing nearly all of the particle density into the outer wells:  the particle density becomes significantly different from zero only in the outer wells where though only single occupation remains non-negligible.  For this case the entanglement would be expected to be bounded from below by $2/M$.  Fig.~\ref{fig:morepointsUdashcompare} shows that in the long range case the entanglement does indeed come close to this limit.  This density configuration is not accessible with the contact-type interaction: this interaction could affect the density profile to some extent, but could not make the outer wells more favourable than the inner wells (see \cite{Coe-Viv-DAm} and fig.~\ref{fig:densitycompare}).  The EHM reproduces fairly well the effect on the density of long-range interactions for $U \lesssim 10$, and we see that the related entanglement saturates at a value lower than the HM for higher values of $U$.  However this saturation value is still much higher than the entanglement values reached by the QD system with long range interactions, which do not yet saturate even for $U$ as large as $150$.  On the other hand, the greater effect on the density distribution means that the QD system with long range interaction comes closer to achieving the theoretical maximum entanglement for $M=4$, $S^{th}_{max}=1.623$ (see table I in ref.~\cite{Coe-Viv-DAm}):
for long-range interactions $S_{\text{max}}=1.565$ compared with $S_{\text{max}}=1.550$ for the contact interaction (see fig.~\ref{fig:4wellscomparedeltatolongrangeandhubbard}).

\begin{figure}[ht]\centering
  \includegraphics[width=.4\textwidth]{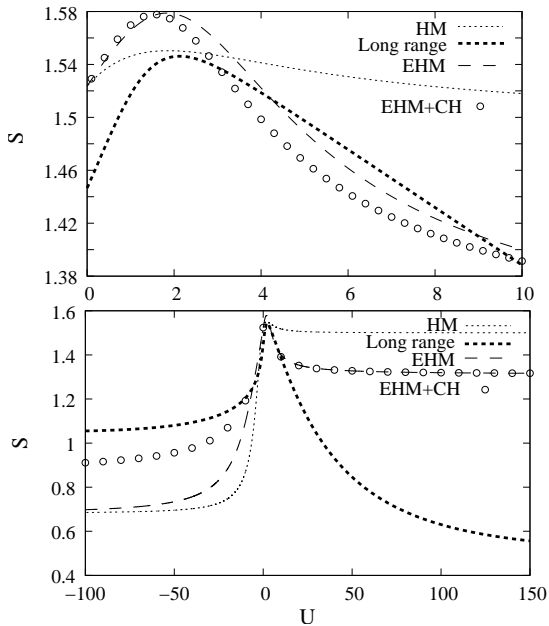}
  \caption{Average single-site entanglement vs interaction strength $U$ for $M=4$.  Upper panel: HM (dotted line); long range interaction,  $d=0$, $w=2$ $a_{0}$, $U=1.84C_{U}$ (thick dashed line); EHM with $U'=0.54U$ (thin dashed line); EHM+CH with $U'=0.54U$, $t'=0.047U$ (circles). For all QD systems $v_0=10$ Hartree. Lower panel: same as upper panel but for a greater range of interaction strengths.} \label{fig:4d=0wellscomparedeltatolongrange}
\end{figure}

We see in fig.~\ref{fig:4d=0wellscomparedeltatolongrange} that in the limiting case of $d=0$, for which our QD model reduces to a single dot of width $Mw$, the EHM is fairly accurate for the entanglement of the QD system when $|U|$ is small. However the maximum and $U\sim0$ values are less accurate than with $d=2$ $a_0$: compare the upper panels of figs.~\ref{fig:morepointsUdashcompare} and~\ref{fig:4d=0wellscomparedeltatolongrange}. We note that for $d=0$ the HM and EHM are also substantially less accurate for $U<0$ than in the $d=2$ $a_{0}$ case (compare lower panels of figs.~\ref{fig:morepointsUdashcompare} and \ref{fig:4d=0wellscomparedeltatolongrange}).  Including the correlated hopping term modifies only slightly the entanglement of the EHM for $U>0$, but has a noticeable effect for $U<0$ where it produces a better approximation to the long range QD system.

The Hubbard model with $U<8$ is used to describe strongly correlated solids \cite{RevModPhys.70.1039}, while larger $U$ (up to $\sim 150$) can represent systems of optically trapped atoms \cite{JordensNature2008}.  To estimate the typical values of $U$ in a quantum dot system,  we may use effective atomic units and the observation that a system with $C_{U}=K$, well depth $v_{0}$, well width $w$, barrier width $d$ and minimum interaction length $l$ is equivalent to a system with $C_{U}=1$, well depth $v_{0}/K^{2}$, well width $Kw$, barrier width $Kd$ and minimum interaction length $Kl$. Under these conditions, given a specific QD system and the well parameters, the on-site $U$ can be estimated, as shown in table \ref{tbl:estimateU}.  We then note that some typical QD parameters such as in \cite{Biolatti} and \cite{NatureSotier} correspond to values of $U$ for which the EHM and EHM+CH are unable to reproduce the QD entanglement. 

The results in table \ref{tbl:estimateU} suggest that for a long range interaction the HM, EHM and EHM+CH are not at all suitable to accurately model the average single-site entanglement of realisable quantum dots when $d$ is comparable or larger than $w$.  However they all appear to be a fair approximation to the average single-site entanglement in the limiting case of $d=0$ and small $U$ suggesting that the HM, EHM and EHM+CH are useful in modelling systems of dots when $d$ is much smaller than $w$ as such systems correspond to small $U$. Interestingly, this may include the case of a single dot when modelled as a (small) set of finite partitions (system GaAs$^{(3)}$ in table \ref{tbl:estimateU}).

\begin{table*}[t!]
\centering
\caption{Estimate of the on-site interaction strength $U$ for different QD systems: GaAs-type systems with reduced mass $m_{eff}=0.067m_e$ and dielectric constant $\epsilon=10.9$ (GaAs$^{(1)}$ to GaAs$^{(3)}$); GaAs-based system with $m_{eff}=0.067m_e$ and $\epsilon=12.1$ (GaAs$^{(4)}$ \cite{Biolatti}); and CdSe-based system with $m_{eff}=0.45m_e$ and $\epsilon=9.1$ \cite{NatureSotier}. For the $d=0$ limiting case, the system is physically a {\it single} dot of width $44$nm but modelled as four partitions of width $w=11$ nm.} \label{tbl:estimateU}
\begin{tabular}{ | l || c  c  c  | c || c c c |c| r|}
\hline
{\bf System} &  \multicolumn{4}{c||}{\bf QD Parameters} &  \multicolumn{5}{c|}{\bf Corresponding Long-range model parameters}   \\
 &  $v_0$(eV) & $w$(nm) & $d$(nm) & $l$ (nm) & $v_0$(Hartree) & $w(a_0)$ & $d(a_0)$ & $l(a_0)$ & $\sim U$  \\
\hline
GaAs$^{(1)}$  & 1.5 & 5.5 & 5.5 & 2.75 & 10 & 2 & 2 &1& $5700$ \\
GaAs$^{(2)}$ & 1.2 & 6.3 & 1.9 & 3.15  & 10 & 2 & 0.6 &1&$10$\\
GaAs$^{(3)}$ & 0.4 & 11.2 & 0 & 5.6 & 10 & 2 & 0 &1& $1.2$ \\
\hline
GaAs$^{(4)}$ \cite{Biolatti}  & 1.0 & 5.0 & 5.0 & 3.4& 10 & 1.48 & 1.48 &1& $500$ \\
\hline
CdSe \cite{NatureSotier}& 0.6 & 2.0 & 2.0 & 1.7 & 10 & 1.19 & 1.19 &1& $1000$ \\
\hline
\end{tabular}
\end{table*}

\section{Particle-particle spatial entanglement}
We now investigate spatial entanglement \cite{Coe}, {\it i.e.} the particle-particle entanglement related to the spatial degrees of freedom.  For the QD system the related reduced density matrix is infinite dimensional and given by  
\begin{equation}
\rho_{\text{red,spatial}}(x,x')=\int \Psi_{\text{QD}}^{\star}(x,x_{\tau}) \Psi_{\text{QD}}(x',x_{\tau}) dx_{\tau}
\end{equation}
where $\Psi_{\text{QD}}$ is the QD system wavefunction. 
The von Neumann entropy is then $S_\text{spatial}=-Tr \rho_\text{red,spatial} \log_{2} \rho_\text{red,spatial}$.
We consider the QD system with both contact and long-range interactions, and compare it with the HM and the EHM (with and without the CH term) with $M$ sites. To calculate the HM (EHM) spatial entanglement we transform the HM (EHM) wavefunction from the site occupation basis to the particle basis to give $\Psi_{\text{HM (EHM)}}$. We move from having a superposition of the tensor products of $M$ sites, where each site can be in one of four states, to a superposition of the tensor products of two particles, each of which can be in one of $M$ sites.  This mapping, for the {\it two}-electron system under consideration, is given by
\ba
\nonumber \ket{0}_{1}\ldots\ket{\uparrow\downarrow}_{i}\ldots\ket{0}_{M} &\rightarrow&  \ket{x_{i,a}}\ket{x_{i,b}} \\   \otimes\frac{1}{\sqrt{2}}\big[\ket{\uparrow_a}\ket{\downarrow_b}-\ket{\downarrow_a}\ket{\uparrow_b}\big],
\ea
for states' components with doubly occupied sites, and by
\ba
\nonumber \ket{0}_{1}\ldots\ket{\uparrow}_{i}\ldots\ket{\downarrow}_{j}\ldots\ket{0}_{M}& \rightarrow &\frac{1}{\sqrt{2}}\big[\ket{x_{i,a}}\ket{x_{j,b}}\ket{\uparrow_a}\ket{\downarrow_b}\\
-\ket{x_{j,a}}\ket{x_{i,b}}\ket{\downarrow_a}\ket{\uparrow_b}\big],
\ea
for states' components with single-occupied sites, where antisymmetry in particle exchange has been imposed. Here $\{x_{i,s}\}$, with $i=1,2\ldots M$, and $s=a,b$, represent the {\it discrete} M possible coordinates (site positions) for particle $s$.
The spin part of the resulting ground-state wavefunction factorises into a singlet, so its entanglement is constant and, as such, not of interest here. The spatial part of the wavefunction $\Psi_{\text{HM (EHM)}}$ 
can now be used to calculate the $M \times M$ reduced density matrix $\rho_{\text{red,spatial}}^{HM (EHM)}=Tr_{\text{A}} \left|\Psi_{\text{HM (EHM)}} \right\rangle \left\langle \Psi_{\text{HM (EHM)}}\right|$ where we trace out the subsystem $A$ corresponding to one of the particles. The reduced density matrix is then used in the evaluation of the von Neumann entropy $S_{\text{spatial}}$.

\begin{figure}[t!]\centering
  \includegraphics[width=.4\textwidth]{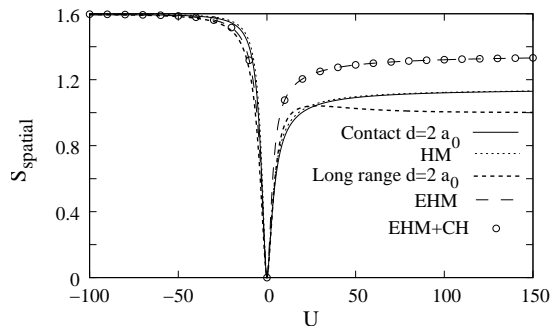} 
  \caption{Spatial entanglement vs interaction strength $U$: contact interaction, $d=w=2$ $a_{0}$, $U=12344C_{U}$ (solid line); HM (dotted line); long-range interaction, $d=w=2$ $a_{0}$, $U=17865C_{U}$ (thick dashed line); EHM (thin dashed line), EHM+CH (circles) with $U'=0.28U$. $v_0=10$ Hartree for all QD systems.}\label {fig:4wellsspatiald=2}
\end{figure}

\begin{figure}[t!]\centering
  \includegraphics[width=.4\textwidth]{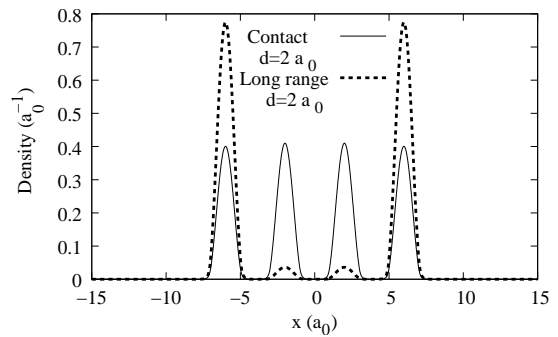}
  \caption{Density profile for the QD system at $U=100$ with $M=4$, $d=w=2$ $a_{0}$, $v_0=10$ Hartree and a contact (solid line) or long range (thick dashed line) interaction.}\label{fig:densitycompare}
\end{figure}

Due to the lattice discretization, as an approximation to $\Psi_{\text{QD}}$, $\Psi_{\text{HM (EHM)}}$ retains very few of the spatial degrees of freedom, {\it as few as 4} in the system at hand.
Yet, our results in fig.~\ref{fig:4wellsspatiald=2} show that the agreement between the EHM spatial and the QD system spatial entanglement is surprisingly good, especially for attractive interactions.  We also find that the HM spatial entanglement is an excellent approximation to the QD system with a contact interaction for both attractive and repulsive interactions (fig.~\ref{fig:4wellsspatiald=2}). Interestingly, the QD system with long range interaction has a lower entanglement for large positive $U$ than the QD system with contact interaction, even if the former has stronger, Coulomb-dependent, correlations. That occurs because the long range interaction shapes the particle density substantially more, which means that for large repulsion almost all of the density resides in the outer wells, see fig.~\ref{fig:densitycompare}.  This results in a triplet-type state~\cite{Shirwan09} for which the spatial entanglement is equivalent to the entanglement of a maximally entangled two-qubit state, {\it i.e.}, unity. Our data confirm this picture. Interestingly in this regime the EHM is actually a poorer approximation than the HM to the long range QD system.  The EHM has increased Coulomb correlations in respect to the HM and hence displays a higher entanglement.  However the EHM fails to effectively exclude the particles from the inner wells and to reproduce the triplet-type state (with its lower entanglement) which characterises the QD system when $U>>10$. In fact at $U=100$ the single-particle occupation at the outer (inner) wells is $0.96$ ($0.045$) for the QD system with long-range interaction, while for the EHM the probability of single-particle occupation at the outer sites is $0.74$ and at the inner sites it is still as high as $0.26$.  This implies that the rank of the EHM density matrix does not reduce to the one of a two-qubit density matrix.

\section{Conclusion}
\label{sec:Conclusions}
We studied the Hubbard, extended Hubbard, and extended Hubbard with correlated hopping models as an approximation to the local and spatial entanglement of a one-dimensional chain of nanostructures, in which trapped particles are interacting via long-range interactions. We focused on a system comprised of $4$ quantum dots, though the typical trends we found apply to longer chains (not shown), as the long-range interaction will eventually force the density into the outer wells.

For well-separated QDs the EHM reproduces fairly well the average-single site entanglement for attractive long-range interactions and $0<U \lesssim 10$, though with a maximum positioned at a weaker interaction strength.  In contrast with the contact interaction case \cite{Coe-Viv-DAm}, for long range interactions and $U \gtrsim 10$ the EHM completely fails to reproduce the single-site entanglement, with an error which rapidly increases up to $\sim150\%$ at $U=500$ (not shown).  We emphasise that $U$-values as high as $5700$ do indeed correspond to some of the typical parameter ranges for QDs, so care must be taken when trying to model nanostructure systems with the HM and EHM. 

We note that ref.~\cite{PEDERSEN07} found the HM  not to be an accurate approximation when considering the exchange coupling in two-electron double quantum-dots modelled with a parabolic confining potential and a Coulomb interaction in two dimensions or an effective model in one dimension.  However the HM did at least qualitatively reproduce the behaviour of the exchange coupling as the inter-dot distance was varied, while, interestingly, the EHM did not.  In contrast we found the EHM to be a better approximation than the HM to the average single-site entanglement of the QD-chain system, though with the limitations described above. 

We considered the limiting case of interdot barrier $d=0$. In this case our model describes a {\it single} QD, which is divided into an arbitrary number $M$ of partitions. We then considered the average entanglement of one of these partitions with the others.   We found that this corresponds to small $U$ for physically realisable dots and in this region the HM and EHM accuracy as an approximation to the entanglement were fairly good. 

We also studied the case of attractive interaction: for well-separated QDs the average single-site entanglement was well approximated by both the HM and the EHM for any interaction strength. However as the interdot barrier width decreases, the HM and EHM fail to reproduce the entanglement of the QD system at medium and large interaction strength, quickly reaching a discrepancy of $\sim 40\%$ for the limiting case $d=0$.

We designed a procedure to calculate the spatial entanglement from the HM and found that the spatial entanglement associated with the HM was a very good approximation to the spatial entanglement of the four QD chain when particle-particle {\it contact} interaction was considered.  This was surprising as we lose most of the spatial degrees of freedom when using the HM to calculate the spatial entanglement: we consider one position per site only -- four positions in total for the specific case analysed in this paper -- compared with an infinite number for any QD system. Yet our results show that the pertinent information seems to be retained. The QD system with long range, repulsive interaction was less well approximated by the HM, as could be expected, but interestingly the EHM performed worse still. However their errors at large $U$, $\sim 13\%$ and $\sim 33 \%$ respectively, were considerably smaller than for the average single-site entanglement of the long-range QD system.  We attributed the worse performance of the EHM to the long range interaction causing most of the electron density to reside in the two outer wells so essentially giving the maximum entanglement of a two-qubit system.  The EHM fails to modify the density to this extent, and hence to lower the entanglement when $U>>10$.  We find that for the analysed QD system the correlated hopping term is very small and produces appreciable effects on the entanglement only in the limiting case of $d=0$ with {\it attractive} interactions.

 Our results show that when dealing with nanostructures care must be taken in assessing the validity of the HM or EHM as an approximation to the system behaviour, as for some of the experimentally relevant parameter range these approximations fail badly to reproduce the local entanglement.  Surprisingly, our calculations show that for QD chains the essential properties of the spatial entanglement are retained when discretizing the spatial degrees of freedom to a handful of points, and in this case the HM and EHM remain fair approximations at all parameter values.

\acknowledgements
JPC and IDA gratefully acknowledge funding from EPSRC grant EP/F016719/1. VVF is supported by Brazilian funding from CAPES(4101/09-0).

\end{document}